\documentclass[english,10pt,twocolumn,twoside]{IEEEtran}
\usepackage[T1]{fontenc}
\usepackage[latin9]{inputenc}
\usepackage{units}
\usepackage{bm}
\usepackage{amsmath}
\usepackage{amssymb}
\usepackage{graphicx}
\usepackage{babel}
\begin{document}

\title{One-bit Decentralized Detection with a Rao Test for Multisensor Fusion}

\author{D.~Ciuonzo,~\IEEEmembership{Student~Member,~IEEE,} G.~Papa,~\IEEEmembership{Student~Member,~IEEE,}
\\
G.~Romano,~\IEEEmembership{Member,~IEEE,} P.~Salvo~Rossi,~\IEEEmembership{Senior Member,~IEEE,}
and P.~Willett,~\IEEEmembership{Fellow,~IEEE}%
\thanks{D. Ciuonzo, G. Papa, G. Romano and P. Salvo Rossi are with the Department
of Industrial and Information Engineering, Second University of Naples,
Aversa (CE), Italy (e-mail: \{domenico.ciuonzo, giuseppe.papa, gianmarco.romano,
pierluigi.salvorossi\}@unina2.it).\protect \\
 P.  Willett is with the Department of Electrical and Computer Engineering,
University of Connecticut, Storrs (CT), US (e-mail: willett@engr.uconn.edu).
\% This work was supported in part by the US Office of Naval Research
under contract N00014-09-10613.%
}\vspace{-0.5cm}
}
\maketitle
\begin{abstract}
In this letter we propose the Rao test as a simpler alternative to
the generalized likelihood ratio test (GLRT) for multisensor fusion.
We consider sensors observing an unknown deterministic parameter with
symmetric and unimodal noise. A decision fusion center (DFC) receives
quantized sensor observations through error-prone binary symmetric
channels and makes a global decision. We analyze the optimal quantizer
thresholds and we study the performance of the Rao test in comparison
to the GLRT. Also, a theoretical comparison is made and asymptotic
performance is derived in a scenario with homogeneous sensors. All
the results are confirmed through simulations.\end{abstract}

\section{Introduction}

\PARstart{D}{ecentralized} detection with wireless sensor networks
(WSNs) has received close attention by the scientific community over
the last decade. Each sensor, rather than sending its observed measurements,
typically sends one bit of information about the estimated hypothesis
to the decision fusion center (DFC), which makes a global decision.
Such an approach is generally employed in order to satisfy stringent
constraints on bandwidth and energy. In this context the optimal test
(under Bayesian and Neyman-Pearson frameworks) at each sensor is well
known to be a one-bit quantization of the local likelihood-ratio test
(LRT). Unfortunately in most cases, due to a lack of signal knowledge,
it is not possible to compute the local LRT at the generic sensor.
Also, even when the sensors \emph{can} compute their local LRT, the
search for local quantization thresholds is well known to be exponentially
complex \cite{Tsitsiklis1993,Viswanathan1997}. In such situations
the raw measurement is directly quantized into a single bit of information;
the DFC is then in charge of solving a composite hypothesis testing
problem.

Some simple approaches have been based on the counting rule or channel-aware
statistics, which neglect the dependence with respect to (w.r.t.)
the unknown signal \cite{Aalo1994,Chen2004,Niu2008,Ciuonzo2012}.
On the other hand, in some particular scenarios the uniformly most
powerful test is independent of the unknown parameters under the alternate
hypothesis, which then do not need to be estimated \cite{Ciuonzo2013}.
Nonetheless, typically the fusion rule employed at the DFC is based
on the generalized LRT (GLRT). GLRT-based fusion of quantized data
was studied in \cite{Niu2006b,Iyengar2012}, for detecting a source
with unknown location and fusing conditionally dependent decisions,
respectively. Recently in \cite{Fang2013} the GLRT has been used
to detect an unknown deterministic signal (in a decentralized fashion
with quantized measurements and noisy communication channels of identical
quality) and an asymptotically optimal threshold choice for the quantizer
has been derived in the non-homogeneous sensor case (i.e. an additive
Gaussian observation model with unequal variances).

The contributions of this letter are summarized hereinafter. We study
the problem in \cite{Fang2013} and we propose the Rao test as a computationally
simpler alternative to the GLRT, since it does not require any estimation
procedure; its closed form is obtained in the more general case of
zero-mean noise with symmetric and unimodal pdf and non-identical
bit-error probabilities (BEPs) on the communication channels. Also, we
discuss the optimal choice of quantizer threshold for some pdfs of
interest. Furthermore, the Rao test is compared to the GLRT through
simulations showing that, in addition to sharing the same asymptotic
distribution, it achieves practically the same performance for a finite
number of sensors. This result becomes in fact theoretical coincidence
in a scenario with homogeneous sensors; for the latter scenario a
tighter asymptotic distribution of both tests is derived. 

The letter is organized as follows: Sec. \ref{sec:Problem statement}
introduces the model; in Sec. \ref{sec:Rao Test} we derive the Rao
test and the corresponding optimal thresholds; in Sec. \ref{sec:GLRT and Rao - homogeneous scenario}
the GLR and Rao tests are compared analytically in a homogeneous scenario,
while in Sec. \ref{sec:Numerical-Results} we confirm the results
through simulations; in Sec. \ref{sec:Conclusions} we draw some conclusions.

\section{Problem statement\label{sec:Problem statement}}

The system model is described%
\footnote{\emph{Notation} - Lower-case bold letters denote vectors, with $a_{n}$
representing the $n$th element of $\bm{a}$; upper-case calligraphic
letters, e.g. $\mathcal{A}$, denote finite sets; $\mathbb{E}\{\cdot\}$,
$\mathrm{var\{\cdot\}}$ and $(\cdot)^{t}$ denote expectation, variance
and transpose, respectively; $P(\cdot)$ and $p(\cdot)$ are used
to denote probability mass functions (pmf) and probability density
functions (pdf), respectively, while $P(\cdot|\cdot)$ and $p(\cdot|\cdot)$
their corresponding conditional counterparts; $\mathcal{N}(\mu,\sigma^{2})$
denotes a normal distribution with mean $\mu$ and variance $\sigma^{2}$;
$\chi_{k}^{2}$ (resp. $\chi_{k}^{'2}(\xi)$) denotes a chi-square
(resp. a non-central chi-square) distribution with $k$ degrees of
freedom (resp. and non-centrality parameter $\xi$); $\mathcal{U}(a,b)$
denotes a continuous-valued uniform distribution with support set
$[a,b]$; $\mathcal{L}(\mu,\beta)$ denotes a Laplace distribution
with mean $\mu$ and scale parameter $\beta$; the symbols $\sim$
and $\overset{a}{\sim}$ mean \textquotedblleft{}distributed as\textquotedblright{}
and ``asymptotically distributed as''.%
} as follows. We consider a binary hypothesis testing problem in which
a collection of sensors $k\in\mathcal{K}\triangleq\{1,\ldots,K\}$
collaborate to detect the presence of an unknown deterministic parameter
$\theta\in\mathbb{R}$. The problem can be summarized as follows:
\begin{gather}
\begin{cases}
\mathcal{H}_{0}\quad:\quad & x_{k}=w_{k},\\
\mathcal{H}_{1}\quad:\quad & x_{k}=h_{k}\theta+w_{k},\qquad k\in\mathcal{K};
\end{cases}\label{eq:binary_test}
\end{gather}
where $x_{k}\in\mathbb{R}$ denotes the $k$th sensor measurement,
$h_{k}\in\mathbb{R}$ is a known observation coefficient and $w_{k}\in\mathbb{R}$
denotes the noise random variable (RV) with \emph{$\mathbb{E}\{w_{k}\}=0$
}and \emph{unimodal symmetric} pdf%
\footnote{Noteworthy examples of such pdfs are the Gaussian, Laplace, Cauchy
and generalized Gaussian  distributions with zero mean \cite{Kay1998}.%
}, denoted with $p_{w_{k}}(\cdot)$. Furthermore, the RVs $w_{k}$
are assumed mutually independent. It is worth noting that Eq. (\ref{eq:binary_test})
refers to a \emph{two-sided test }\cite{Kay1998}, where $\{\mathcal{H}_{0},\mathcal{H}_{1}\}$
corresponds to $\{\theta=\theta_{0},\theta\neq\theta_{0}\}$ (in our
case $\theta_{0}=0$). 

Also, to meet stringent bandwidth and power budgets in WSNs, the $k$th
sensor quantizes%
\footnote{We restrict our attention to deterministic quantizers for simplicity;
an alternative is the use of stochastic quantizers, however their
analysis falls beyond the scope of this letter.%
} $x_{k}$ into one bit of information, that is $b_{k}\triangleq u(x_{k}-\tau_{k})$,
$k\in\mathcal{K}$, with $u(\cdot)$ denoting the Heaviside (unit)
step function and $\tau_{k}$ the quantizer threshold. The quantized
measurement $b_{k}$ is sent over a binary symmetric channel (BSC)
and the DFC observes a (communication) error-prone $y_{k}$, that
is $y_{k}=b_{k}$ with probability $1-P_{e,k}$ and $y_{k}=1-b_{k}$
with probability $P_{e,k}$, which we collect as $\bm{y}\triangleq\left[\begin{array}{ccc}
y_{1} & \cdots & y_{K}\end{array}\right]^{t}$. Here $P_{e,k}$ denotes the BEP of $k$th link. The problem here
is the derivation of a (computationally) simple test on the basis
of $\bm{y}$ and the quantizer design for each sensor (i.e. an optimized
$\tau_{k}$, $k\in\mathcal{K}$).

\section{Rao Test \label{sec:Rao Test}}

\subsection{Test derivation}

A common approach to detection in composite hypothesis testing problems
is given by the GLRT, which has been derived and studied in \cite{Fang2013}
for the model under investigation and whose expression is:
\begin{equation}
\left\{ \Lambda_{\mathrm{G}}\triangleq2\cdot\ln\left[\frac{P(\bm{y};\hat{\theta}_{1})}{P(\bm{y};\theta_{0})}\right]\right\} \begin{array}{c}
{\scriptstyle \hat{\mathcal{H}}=\mathcal{H}_{1}}\\
\gtrless\\
{\scriptstyle \hat{\mathcal{H}}=\mathcal{H}}_{0}
\end{array}\gamma\label{eq:GLRT_general}
\end{equation}
where $P(\bm{y};\theta)$ denotes the likelihood as a function of
$\theta$, $\hat{\theta}_{1}$ is the \emph{maximum likelihood }(ML)\emph{
estimate} under $\mathcal{H}_{1}$ (i.e. $\hat{\theta}_{1}\triangleq\arg\max_{\theta}P(\bm{y};\theta)$)
and $\gamma$ is the threshold. It is clear from Eq. (\ref{eq:GLRT_general})
that $\Lambda_{\mathrm{G}}$ requires the solution to an optimization
problem; this increases the computational complexity of its implementation.
However, in the special case $w_{k}\sim\mathcal{N}(0,\sigma_{k}^{2})$
it was shown in \cite{Ribeiro2006a} that ML estimation is a convex
problem and thus it can be efficiently solved with local-optimization
routines. Unfortunately a closed form for $\hat{\theta}_{1}$ is not
available even under such an assumption.

As such, we pursue the derivation of the Rao test \cite{Kay1998},
which for the scalar case ($\theta\in\mathbb{R}$) is given implicitly
in the form:
\begin{equation}
\left\{ \Lambda_{\mathrm{R}}\triangleq\nicefrac{\left(\left.\frac{\partial\ln P(\bm{y};\theta)}{\partial\theta}\right|_{\theta=\theta_{0}}\right)^{2}}{\mathrm{I}(\theta_{0})}\right\} \begin{array}{c}
{\scriptstyle \hat{\mathcal{H}}=\mathcal{H}_{1}}\\
\gtrless\\
{\scriptstyle \hat{\mathcal{H}}=\mathcal{H}}_{0}
\end{array}\gamma\label{eq:Rao_general}
\end{equation}
where $\mathrm{I}(\theta_{0})$ is the \emph{Fisher information }(FI),
i.e. $\mathrm{I}(\theta)\triangleq\mathbb{E}\{\left(\frac{\partial\ln\left[P(\bm{y};\theta)\right]}{\partial\theta}\right)^{2}\}$
evaluated at $\theta_{0}$. The motivation of our choice is  the extreme
simplicity of the test implementation (since $\hat{\theta}_{1}$ is
not required, cf. Eq. (\ref{eq:Rao_general})), but with the same
weak-signal asymptotic performance as the GLRT, as supported from
the theory \cite{Kay1998}.

In order to obtain $\Lambda_{\mathrm{R}}$ explicitly, we expand $\ln\left[P(\bm{y};\theta)\right]$
as:
\begin{gather}
\ln\left[P(\bm{y};\theta)\right]=\sum_{k=1}^{K}\ln\left[P(y_{k};\theta)\right]=\nonumber \\
\sum_{k=1}^{K}\{y_{k}\cdot\ln\left[(1-P_{e,k})\alpha_{k}(\theta)+P_{e,k}(1-\alpha_{k}(\theta))\right]+\nonumber \\
(1-y_{k})\cdot\ln\left[(1-P_{e,k})(1-\alpha_{k}(\theta))+P_{e,k}\alpha_{k}(\theta)\right]\}\label{eq:likelihood_error_prone}
\end{gather}
where $\alpha_{k}(\theta)\triangleq F_{w_{k}}(\tau_{k}-h_{k}\theta)$,
with $F_{w_{k}}(\cdot)$ denoting the complementary cumulative distribution
function of $w_{k}$. On the other hand, $\mathrm{I}(\theta)$ is
given in closed form \cite{Fang2013} as:
\begin{align}
\mathrm{I}(\theta)= & \sum_{k=1}^{K}\left\{ \frac{(1-2\, P_{e,k})^{2}\cdot h_{k}^{2}\cdot p_{w_{k}}^{2}(\tau_{k}-h_{k}\theta)}{P_{e,k}+(1-2\, P_{e,k})\cdot F_{w_{k}}(\tau_{k}-h_{k}\theta)}\times\right.\nonumber \\
 & \left.\frac{1}{\left[1-P_{e,k}-(1-2\, P_{e,k})\cdot F_{w_{k}}(\tau_{k}-h_{k}\theta)\right]}\right\} .\label{eq:FI_error_prone}
\end{align}
Combining Eqs. (\ref{eq:likelihood_error_prone}) and (\ref{eq:FI_error_prone})
we obtain $\Lambda_{\mathrm{R}}$ in closed form, as shown in Eq.
(\ref{eq:Rao Test explicit- BSC case}) at the top of next page. It
is apparent that $\Lambda_{\mathrm{R}}$ (as well as $\Lambda_{\mathrm{G}}$)
is a function of $\tau_{k}$, $k\in\mathcal{K}$, which can be optimized
in order to achieve (asymptotically) optimal performance. 
\begin{figure*}[t]
\begin{equation}
\Lambda_{\mathrm{R}}={\displaystyle \left(\sum_{k=1}^{K}\frac{(1-2\cdot P_{e,k})\cdot h_{k}\cdot p_{w_{k}}(\tau_{k})\cdot\left[2y_{k}-1\right]}{(1-P_{e,k})\cdot F_{w_{k}}(\tau_{k})^{y_{k}}\cdot\left[1-F_{w_{k}}(\tau_{k})\right]^{1-y_{k}}+P_{e,k}\cdot F_{w_{k}}(\tau_{k})^{1-y_{k}}\cdot\left[1-F_{w_{k}}(\tau_{k})\right]^{y_{k}}}\right)^{2}}\times\left(\mathrm{I}(\theta=0)\right)^{-1}\label{eq:Rao Test explicit- BSC case}
\end{equation}

\hrulefill 
\vspace*{0pt}
\end{figure*}

\subsection{Quantizer design with asymptotic performance analysis}

\begin{figure}
\centering{}\includegraphics[width=0.95\columnwidth]{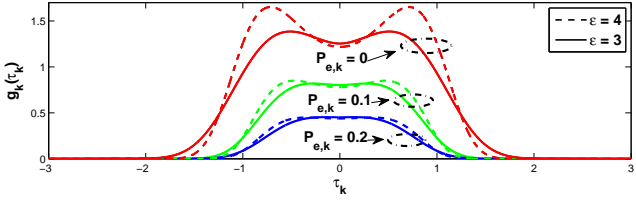}\caption{Effect of $P_{e,k}$ on $g_{k}(\tau_{k})$ when $p_{w_{k}}(\tau_{k})=\frac{\epsilon}{2\alpha\Gamma(1/\epsilon)}\exp\left[-\left(\frac{|\tau_{k}|}{\alpha}\right)^{\epsilon}\right]$;
$\alpha=1$, $\epsilon\in\{3,4\}$ and $P_{e,k}\in\{0,0.1,0.2\}$.\label{fig:bimodality_gtau}}
\end{figure}

We know from theory that $\Lambda_{\mathrm{R}}$ (as well as $\Lambda_{\mathrm{G}}$),
is asymptotically (when the signal is weak%
\footnote{ That is $|\theta_{1}-\theta_{0}|=c/\sqrt{K}$ for some constant $c>0$
\cite{Kay1998}.%
}) distributed as follows \cite{Kay1998}:
\begin{equation}
\Lambda_{\mathrm{R}}\overset{{\scriptstyle a}}{\sim}\begin{cases}
\chi_{1}^{2} & \qquad\mathrm{under}\quad\mathcal{H}_{0}\\
\chi_{1}^{'2}(\lambda_{Q}) & \qquad\mathrm{under}\quad\mathcal{H}_{1}
\end{cases}\label{eq:Asymptotic_performance}
\end{equation}
where the non-centrality parameter $\lambda_{Q}$ is given by:
\begin{equation}
\lambda_{Q}\triangleq(\theta_{1}-\theta_{0})^{2}\mathrm{I}(\theta_{0})
\end{equation}
 with $\theta_{1}$ being the true value under $\mathcal{H}_{1}$.
Clearly the larger $\lambda_{Q}$, the better the GLRT and Rao tests
will perform. Also, as shown in \cite{Fang2013}, $\lambda_{Q}$ is
a function of $\tau_{k}$, $k\in\mathcal{K}$; therefore we choose
$\tau_{k}$, $k\in\mathcal{K}$, in order to maximize $\lambda_{Q}$,
that is

\begin{align}
\arg\max_{\{\tau_{1},\ldots,\tau_{K}\}} & \left\{ \lambda_{Q}=\theta^{2}\sum_{k=1}^{K}\left[\frac{(1-2\, P_{e,k})^{2}\cdot h_{k}^{2}\cdot p_{w_{k}}^{2}(\tau_{k})}{P_{e,k}+(1-2\, P_{e,k})\cdot F_{w_{k}}(\tau_{k})}\times\right.\right.\nonumber \\
 & \left.\left.\frac{1}{1-P_{e,k}-(1-2\, P_{e,k})\cdot F_{w_{k}}(\tau_{k})}\right]\right\} ,\label{eq:non_centrality_parameter_error_prone}
\end{align}
which can be decoupled into the following set of $K$ independent
threshold design problems:
\begin{equation}
\arg\max_{\tau_{k}}\left\{ g_{k}(\tau_{k})\triangleq\frac{p_{w_{k}}^{2}(\tau_{k})}{\Delta_{k}+F_{w_{k}}(\tau_{k})\cdot\left[1-F_{w_{k}}(\tau_{k})\right]}\right\} \label{eq:objective_function_bsc}
\end{equation}
where $\Delta_{k}\triangleq[P_{e,k}\cdot(1-P_{e,k})]/(1-2P_{e,k})^{2}$.
It is known from quantized estimation literature \cite{Papadopoulos2001,Rousseau2003} that
many unimodal and symmetric $p_{w_{k}}(\cdot)$'s with $\mathbb{E}\{w_{k}\}=0$
lead to $\tau_{k}^{*}\triangleq\arg\max_{\tau_{k}}g_{k}(\tau_{k})=0$
(independent of $\Delta_{k}$); such examples are the Gaussian, Laplace,
Cauchy and the widely used generalized normal distribution, that is
$p_{w_{k}}(\tau_{k})=\frac{\epsilon}{2\alpha\Gamma(1/\epsilon)}\exp\left[-\left(\frac{|\tau_{k}|}{\alpha}\right)^{\epsilon}\right]$,
only when $0\leq\epsilon\leq2$; on the other hand when $\epsilon>2$,
$g_{k}(\tau_{k})$ becomes bimodal (since it is symmetric) as shown
in Fig. \ref{fig:bimodality_gtau}.  However the effect of a non-ideal
BSC smoothes the gain achieved by $\tau_{k}^{*}$ and thus $\tau_{k}=0$
is still a good (sub-optimal) choice.
 Substituting $\tau_{k}=0$, $k\in\mathcal{K},$ in Eq.~(\ref{eq:Rao Test explicit- BSC case}),
leads to the following simplified expression for threshold-optimized
Rao test (denoted with $\Lambda_{\mathrm{R}}^{*}$):
\begin{equation}
\Lambda_{\mathrm{R}}^{*}=\frac{4\cdot\left[\sum_{k=1}^{K}(1-2\, P_{e,k})\cdot p_{w_{k}}(0)\cdot h_{k}\cdot(y_{k}-\frac{1}{2})\right]^{2}}{\sum_{k=1}^{K}(1-2\, P_{e,k})^{2}\cdot p_{w_{k}}^{2}(0)\cdot h_{k}^{2}}\label{eq:Rao_test_simplified_error_prone}
\end{equation}
which is considerably simpler than the GLRT, as it obviates solution
of an optimization problem (which depends on $p_{w_{k}}(\cdot)$).
Furthermore, the corresponding optimized non-centrality parameter,
denoted with $\lambda_{Q}^{*}$, is given by: 
\begin{equation}
\lambda_{Q}^{*}=4\,\theta^{2}\cdot\sum_{k=1}^{K}\left[(1-2\, P_{e,k})^{2}\cdot p_{w_{k}}^{2}(0)\cdot h_{k}^{2}\right]\label{eq:optimized-noncentralitypar-BSC-1}
\end{equation}
\emph{Remarks} - In the case of BSCs of the same quality (i.e. $P_{e,k}=P_{e},$
$k\in\mathcal{K}$) we simply get $\lambda_{Q}^{*}=(1-2\, P_{e})^{2}\cdot\lambda_{Q_{0}}^{*},$
where $\lambda_{Q_{0}}^{*}\triangleq4\,\theta^{2}\cdot\sum_{k=1}^{K}\left[p_{w_{k}}^{2}(0)\cdot h_{k}^{2}\right]$
represents $\lambda_{Q}^{*}$ in the ideal BSC case ($P_{e,k}=0$,
$k\in\mathcal{K}$). This result generalizes the one in \cite{Fang2013},
by stating that the \emph{loss due to non-ideal communications is
asymptotically independent} \emph{of} $p_{w_{k}}(\cdot)$, $k\in\mathcal{K}$.

\section{Comparison in homogeneous scenario\label{sec:GLRT and Rao - homogeneous scenario}}

In this section we study the simplified scenario $h_{k}=h$, $p_{w_{k}}(\cdot)=p_{w}(\cdot)$,
$P_{e,k}=P_{e}$, $k\in\mathcal{K}$, to get an intuitive interpretation
of the two threshold-optimized tests ($\tau_{k}^{*}=0$). Based on
these assumptions, the statistics in Eqs. (\ref{eq:GLRT_general})
and (\ref{eq:Rao_test_simplified_error_prone}) reduce to:
\begin{eqnarray}
\Lambda_{\mathrm{G}}^{*} & = & 2K\cdot\left[\hat{\rho}\ln\left(\frac{\hat{\rho}}{\rho_{0}}\right)+(1-\hat{\rho})\ln\left(\frac{1-\hat{\rho}}{1-\rho_{0}}\right)\right]\label{eq:GLRT_KL_pre}\\
 & = & 2K\cdot D_{\mathrm{KL}}(\hat{P}(y_{k})\parallel P(y_{k};\theta_{0}))\label{eq:GLRT_hoeffding_KL}\\
\Lambda_{\mathrm{R}}^{*} & = & 4K\cdot\left[\hat{\rho}-\rho_{0}\right]^{2}\label{eq:Rao_TVD_pre}\\
 & = & 4K\cdot\left[D_{\mathrm{TVD}}(\hat{P}(y_{k})\parallel P(y_{k};\theta_{0}))\right]^{2}\label{eq:Rao_TVD}
\end{eqnarray}
where $\Lambda_{\mathrm{G}}^{*}\triangleq\Lambda_{\mathrm{G}}(\tau_{k}=0)$,
$\hat{\rho}\triangleq\sum_{k=1}^{K}y_{k}/K$ and $\rho_{0}\triangleq\nicefrac{1}{2}$.
Here $\hat{P}(y_{k})$ represents the empirical distribution of the
i.i.d. binary source $\{y_{1},\ldots,y_{K}\}$ and $D_{\mathrm{KL}}(\cdot\parallel\cdot)$
and $D_{\mathrm{TVD}}(\cdot\parallel\cdot)$ denote the \emph{Kullback-Leibler
}(KL) and \emph{total variation distance }(TVD) divergences, respectively
\cite{Cover2006}. It is worth noticing that in Eq. (\ref{eq:GLRT_hoeffding_KL})
we exploited the closed form of $\hat{\theta}_{1}=-\frac{1}{h}F_{w}^{-1}\left(\left(\hat{\rho}-P_{e}\right)/\left(1-2P_{e}\right)\right)$
(see \cite{Ribeiro2006a} for a similar result).

Exploiting KL%
\footnote{Since it is increasing when $\hat{\rho}>\rho_{0}$ and symmetric around
$\rho_{0}$.%
} and TVD divergences properties it can be shown that both Eqs. (\ref{eq:GLRT_hoeffding_KL})
and (\ref{eq:Rao_TVD}) are monotone (increasing) functions of $|\hat{\rho}-\rho_{0}|$
and therefore  \emph{represent equivalent tests} in a homogeneous
sensor scenario, meaning their performances coincide also for a finite
number of sensors.

Finally, we derive a tighter asymptotic form of the conditional pdf
(\emph{not requiring} the weak-signal assumption) of both the tests
in this scenario with the help of the central limit theorem (CLT)
\cite{Cover2006}. Without loss of generality we focus hereinafter
on $\Lambda_{\mathrm{R}}^{*}$ (since $\Lambda_{\mathrm{G}}^{*}$
has the same performance). For this purpose, we define the RV $\xi\triangleq\frac{\sum_{k=1}^{K}\left(2y_{k}-1\right)}{\sqrt{K}}$
and we consider the asymptotic form of $p_{\xi}(\cdot|\mathcal{H}_{i})$,
$i\in\{0,1\}$, which according to the CLT is given as $K\rightarrow+\infty$
by:
\begin{gather}
\xi|\mathcal{H}_{0}\;\overset{{\scriptstyle a}}{\sim}\;\mathcal{N}(0,1)\qquad\quad\xi|\mathcal{H}_{1}\;\overset{{\scriptstyle a}}{\sim}\;\mathcal{N}(\sqrt{K}\tilde{\mu}_{1},\tilde{\sigma}_{1}^{2})\label{eq:CLT-based performance}
\end{gather}
where $\tilde{\mu}_{1}\triangleq(1-2P_{e})(2\rho_{1}-1)$ , $\tilde{\sigma}_{1}^{2}\triangleq4\cdot[1+P_{e}(2\rho_{1}-1)-\rho_{1}]\cdot[\rho_{1}+(1-2\rho_{1})P_{e}]$
and $\rho_{1}\triangleq F_{w}(-h\theta)$. From inspection of Eq.~(\ref{eq:Rao_TVD_pre}),
it can be readily verified that  $\Lambda_{\mathrm{R}}^{*}=\xi^{2}$
holds, which can be exploited to obtain closed form performance expressions.

\section{Numerical Results\label{sec:Numerical-Results}}

In this section we compare the Rao test to the GLRT. We evaluate the
performance in terms of system false alarm and detection probabilities,
defined as $P_{F_{0}}\triangleq\Pr\{\Lambda>\gamma|\mathcal{H}_{0}\}$
and $P_{D_{0}}\triangleq\Pr\{\Lambda>\gamma|\mathcal{H}_{1}\}$, respectively,
where $\Lambda$ is the statistic employed at the DFC. We also define
the $k$th sensor observation signal-to-noise ratio (SNR) as $\Gamma_{k}\triangleq\left(h_{k}^{2}\theta^{2}/\mathbb{E}\{w_{k}^{2}\}\right)$.

\begin{figure}[t]
\centering{}\includegraphics[width=0.94\columnwidth]{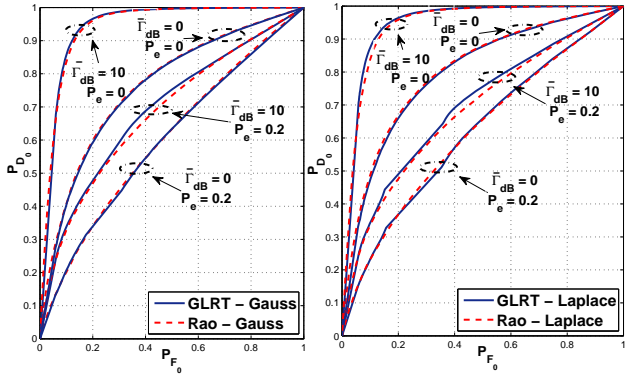}\caption{$P_{D_{0}}$ vs $P_{F_{0}}$; WSN with $K=5$ sensors, $h_{k}\sim\mathcal{U}(0,a)$,
$\theta=1$, $\mathbb{E}\{w_{k}^{2}\}=1$ for Gaussian and Laplace
noise;  $P_{e}\in\{0,0.2\}$,  $\bar{\Gamma}_{dB}\in\{0,10\}$.\label{fig: ROC}}
\end{figure}

In Fig. \ref{fig: ROC} we illustrate $P_{D_{0}}$ vs $P_{F_{0}}$
in a WSN with $K=5$ sensors where $\theta=1$, $h_{k}\sim U(0,a)$,
$k\in\mathcal{K}$ (but known at the DFC), and two noise pdfs: ($i$)
$w_{k}\sim\mathcal{N}(0,\sigma_{k}^{2})$ and $(ii)$ $w_{k}\sim\mathcal{L}(0,\beta_{k})$,
such that $\mathbb{E}\{w_{k}^{2}\}=1$. We consider four combinations
corresponding to $P_{e,k}=P_{e}\in\{0,0.2\}$ and $\bar{\Gamma}_{dB}\in\{0,10\}$,
where we have denoted $\bar{\Gamma}\triangleq\mathbb{E}\{\Gamma_{k}\}$
(in our case $\bar{\Gamma}=(a^{2}\theta^{2})/3\cdot\mathbb{E}\{w_{k}^{2}\}$)
as the average observation SNR. The figures are based on $10^{5}$
Monte Carlo runs. First, it is apparent that the performances of the
GLR and the Rao tests are practically the same for all the considered
scenarios; however the implementation of the Rao test is much simpler
than that of the GLRT. Also, the difference in performances under
Laplacian and Gaussian noises is significant only at $\bar{\Gamma}_{dB}=0$,
while at $\bar{\Gamma}_{dB}=10$ the curves almost overlap. This is
explained since when $\bar{\Gamma}$ is low the  signal  is more concentrated
around zero. Then the imbalance  in the binary pmf observed at the
output of each quantizer is higher when $w_{k}\sim\mathcal{L}(0,\beta_{k})$.

\begin{figure}[t]
\centering{}\includegraphics[width=0.88\columnwidth]{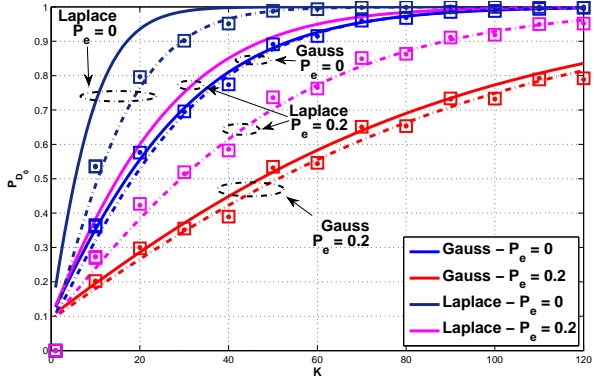}\caption{$P_{D_{0}}$ vs $K$; $P_{F_{0}}=0.1$. Setup: $\theta=0.5$; $h_{k}=1$,
$\mathbb{E}\{w_{k}^{2}\}=1$ ($(\Gamma_{k})_{dB}\approx-6$), $P_{e,k}\in\{0,0.2\}$,
$k\in\mathcal{K}$ (homogeneous scenario). Square ($\square$) and
bullet ($\bullet)$ markers refer to GLRT and Rao test, respectively;
solid and dash-dot lines refer to weak-signal and CLT-based asymptotic
pdfs, respectively. \label{fig: Pd0 vs K}}
\end{figure}

In Fig. \ref{fig: Pd0 vs K} we show $P_{D_{0}}$ as a function of
$K$, assuming $P_{F_{0}}=0.1$. We consider $\theta=0.5$, $h_{k}=1$
and two noise pdfs: ($i$) $w_{k}\sim\mathcal{N}(0,\sigma_{k}^{2})$
and $(ii)$ $w_{k}\sim\mathcal{L}(0,\beta_{k})$, such that $\mathbb{E}\{w_{k}^{2}\}=1$
(thus $(\Gamma_{k})_{dB}\approx-6$), $k\in\mathcal{K}$. Also, we
consider $P_{e}\in\{0,0.2\}$, thus determining a homogeneous scenario.
First, Monte Carlo simulations confirm the theoretical coincidence
between the Rao test (bullet markers) and the GLRT (square markers).
Secondly, it is apparent that the CLT-based performance expressions
(dash-dot) are as accurate as those based on the weak-signal assumption
(solid lines) for Gaussian noise, while in the Laplacian Case the
weak-signal distribution is far from being representative of the distribution.
Interestingly, when $P_{e}=0$, $\lambda_{Q_{0}}^{*}$ in the Laplacian
case coincides with the non-centrality parameter achieved by a GLRT
(or Rao test) based on the raw $x_{k}$, $k\in\mathcal{K}$, given
by $\lambda_{UQ}=\theta^{2}\sum_{k=1}^{K}\frac{h_{k}^{2}}{\beta_{k}^{2}}$,
that is Eq. (\ref{eq:optimized-noncentralitypar-BSC-1}) \emph{does
not predict the loss due to quantization}. On the other hand, by exploiting
the CLT-based performance in Eq. (\ref{eq:CLT-based performance}),
we can compare $\lambda_{UQ}$ with the modified deflection coefficient
of the asymptotic problem given by Eq. (\ref{eq:CLT-based performance})
$d_{Q}\triangleq K\tilde{\mu}_{1}^{2}/\tilde{\sigma}_{1}^{2}$,  
which for the Laplacian noise is given by $d_{Q}=\frac{K\theta^{2}(h^{2}/\beta^{2})}{[1-\exp(-|h\theta|/\beta)]^{2}}\cdot[1-(1-\exp(-\frac{|h\theta|}{\beta})^{2}]$;
thus for this problem we have $\lambda_{UQ}/d_{Q}\approx1.45$, which
predicts the performance loss well. 

\section{Conclusions\label{sec:Conclusions}}

We studied the Rao test for decentralized detection with an unknown
deterministic signal as an attractive alternative to GLRT for a general
model with quantized measurements, zero-mean, unimodal and symmetric
noise (pdf), non-ideal and non-identical BSCs. The asymptotically
optimal sensor thresholds were shown to be zero for many pdfs of interest
and a fair choice in other scenarios; this result was exploited to
 simplify further the Rao test formula. Furthermore, it was shown
through simulations that the Rao test, in addition to being asymptotically
equivalent to the GLRT, achieves practically the same performance
in the finite number of sensors case; for the case of homogeneous
sensors a theoretical coincidence of the two tests was established.
In such a scenario a general asymptotic performance were derived based
on the CLT and not requiring the weak-signal assumption. These latter
were shown to be crucial in performance analysis with peaked noise
pdfs.

\bibliographystyle{IEEEtran}
\bibliography{IEEEabrv,sensor_networks}

\end{document}